\documentclass[amsmath,amssymb,aps,eqsecnum]{revtex4}
\usepackage{epsfig}
\usepackage{changebar}
\input{epsf.tex}
\newcommand{\ket}[1]{|#1\rangle}

\newcommand{\brkt}[2]{\langle#1|#2\rangle}

\begin{document}
\title{Negativity of the Wigner function as an indicator of nonclassicality}
\author{Anatole Kenfack$^1$ and Karol {\.Z}yczkowski$^{2,3}$}

\affiliation{$^1$Max Planck Institute for the physics of complex systems
  N\"othnitzerstr. 38,
  01187 Dresden, Germany}

\affiliation {$^2$ Instytut Fizyki im. Smoluchowskiego,
Uniwersytet Jagiello{\'n}ski,
ul. Reymonta 4, 30-059 Krak{\'o}w, Poland}

\affiliation{$^3$Centrum Fizyki Teoretycznej, Polska Akademia Nauk,
Al. Lotnik{\'o}w 32/44, 02-668 Warszawa, Poland}

\date{\today}

\begin{abstract}
A measure of nonclassicality of quantum states based on the
volume of the negative part of the Wigner function is proposed.
We analyze this quantity for Fock states, squeezed displaced Fock states and
cat-like states defined as coherent superposition of two Gaussian wave packets.
\end{abstract}
\maketitle
\medskip
\begin{center}
{\small e-mail: kenfack@mpipks-dresden.mpg.de
  \ \quad \ karol@cft.edu.pl}
\end{center}

\section{Introduction}
Analyzing pure quantum states in an infinite dimensional Hilbert space it is
useful
 to distinguish a family of {\sl coherent states}, localized in the classical
phase space
and minimizing the uncertainty principle.
These quantum analogues of points in the classical phase space are often
considered
as 'classical' states. For an arbitrary quantum state one may pose a
natural question, to what extent it is 'nonclassical' in a sense that
its properties differ from that of coherent states. In other words, is
there any parameter that may legitimately reflect the degree of
nonclassicality of a given quantum state?
This question was motivated with the first observation of nonclassical
features of electromagnetic fields such as sub-poissonian statistics,
antibunching and squeezing. Additionally, it is well known that
the interaction of (non)linear devices with quantum states may flip
from one state to another; for instance, nonlinear devices may produce
nonclassical states from their interaction with the vacuum or a
classical field.
 A systematic survey of nonclassical properties of
quantum states would be worthwhile because of
the nowadays ever increasing number of
experiments in
nonlinear optics.
An earlier attempt to sheding some light on the nonclassicality of a
quantum state was pioneered by Mandel \cite{Ma79},
who investigated the radiation fields and introduced a parameter ${\bf q}$ measuring
the deviation of the photon number statistics from the Poissonian distribution,
characteristic of coherent states.

In general, to define a measure of nonclassicality of quantum states
one can follow several different approaches \cite{Do02}.
Distinguishing a certain set $\cal C$  of states
(e.g. the set of coherent states $|\alpha\rangle$), one looks for the distance of an analyzed
pure
state $|\psi\rangle$ to this set, by minimizing a distance
$d(|\psi\rangle,|\alpha\rangle)$ over the entire set $\cal C$. Such a scheme
based on the trace distance was first used by Hillery \cite{Hi87,Hi89},
while other distances
(Hilbert-Schmidt distance \cite{DMMW00,DR03} or Bures distance
 \cite{MMS02,MMS03})
 were later used for this purpose.
The same approach is also applicable to characterize mixed quantum states:
minimizing the distance of the density $\rho$ to the set of coherent states is related
\cite{DR03,ABM03}
to  the search for the maximal fidelity
(the Hilbert-Schmidt fidelity ${\rm Tr}\left(\rho\sigma\right)$ or the
Bures-Uhlmann fidelity
$\left({\rm Tr}\sqrt{\rho^{1/2}\sigma \rho^{1/2}}\right)^2$)
with respect to any coherent state, $\sigma=|\alpha\rangle \langle
\alpha|$. On the same footing, the Monge
distance introduced in \cite{zs98,zs01} may be applied to describe, to
what extent a given mixed state is close to the manifold of coherent states.

Yet another way of proceeding is based on the generalized (Cahill) phase space
representation
$R_{\tau}$ of a pure state,
 which interpolates between the Husimi ($Q$), the Wigner ($W$)  and
the Glauber--Sudarshan ($P$)
representations. The Cahill parameter $\tau$
is proportional to the variance of a Gaussian function
one needs to convolute with $P$ representation to obtain $R_{\tau}$
\cite{CG69}. In particular for $\tau=1, 1/2, 0$ one obtains the Q-, W-
and P- representations, respectively.
By construction the $Q$ representation is non-negative for all states,
while the Wigner function may admit also negative values, and the $P$ representation
may be singular or may not exist.

The smoothing effect of $R_{\tau}$
is enhanced as $\tau$ increases. If $\tau$ is large enough so that
$R_{\tau}$ becomes positive definite regular function, thus acceptable
as a classical distribution function, then the smoothing is said to be
complete.
The greatest lower bound $\tau_m$ for the critical value was adopted by Lee
\cite{Le91,Le92}, as {\sl nonclassical depth}
 of a quantum state
 and this approach was further developed in \cite{LB95,MBGB01,MB03}.
The limiting value, $\tau_m=1$, corresponds to the $Q$ function which is always
acceptable as a classical distribution function. The lowest value,
$\tau_m=0$, is ascribed to an arbitrary coherent state because its $P$
function is a Dirac delta function, so its
$\epsilon$--smoothing becomes regular. The range of $\tau_m$ is thus
$\tau_m \in [0,1]$.

If the Husimi function of a pure state admits at least one zero
$Q(\alpha_0)=0$, then a Cahill $R_{\tau}$  distribution  with
a narrower smearing, $\tau<1$, becomes negative in the vicinity of $\alpha_0$.
Therefore the classical depth for such quantum states is maximal, $\tau_m=1$
\cite{LB95}. The only class of states, for which $Q$ representation
has no zeros, are the squeezed vacuum states, for which
$\tau_m$ is a function of the squeezing parameter $s$.
In the limiting case $s=0$ one obtains the coherent state
for which  the $R_0=P$ distribution is a  Dirac delta function,
that is $\tau_m=0$.

A closely related approach to characterizing quantum states
is based on properties of their  Wigner functions in phase space $\{p,q\}$.
One can prove that the Wigner function is bounded from below and from above \cite{CG69}.
In the normalization $\iint W(q,p) dqdp =1$ used later in this work, such a bound reads
$|W(q,p)|\le 1/\pi\hbar$.
Further bound on integrals of the Wigner function were derived
in \cite{BDW99}, while an entropy approach to the Wigner function
was developed in \cite{MF00,Wl01}.

In order to interpret the Wigner function as a classical probability distribution
one needs to require that $W$ is non--negative. As found by Hudson in 1974 \cite{Hu74},
 this is the case for coherent or squeezed vacuum states only.
A possible measure of nonclassicality may
thus be based on the negativity of the Wigner function
which may be interpreted as a signature of quantum interference.

The negativity of the Wigner function has been
linked to nonlocality, according to the Bell inequality \cite{bell87},
 while investigating the original Einstein-Podolsky-Rosen (EPR)
state \cite{einst35}.
In fact Bell argued that the EPR state will
not exhibit nonlocal effects because its Wigner function is everywhere
positive, and as such will allow for a hidden variable description of
correlations. However, it is now demonstrated
\cite{banas98,cohen97} that the Wigner function
of the EPR state, though positive definite, provides a direct evidence
of nonlocality. This violation of the Bell's inequality
holds true for the regularized EPR state \cite{banas99a}
and also for a correlated two-mode quantum state of light \cite{banas99b}.

It is also worth recalling that the Wigner function
can be measured experimentally \cite{SXX93},
including the measurements of its negative values \cite{NRO00}.
The interest put on such experiments has
triggered a search for operational definitions of the Wigner
functions, based on experimental setup \cite{Le97,Lou03}.

The aim of this letter is to study a simple indicator of
the nonclassicality, which depends on the {\sl volume}
of the negative part of the Wigner function.
To demonstrate a potential use of such an approach we
investigate certain families of quantum states.
The nonclassicality indicator is defined in section 2.
The Schr{\"o}dinger cat state, being constructed as coherent
superposition of two Gaussian wave packets,
is analyzed in section 3 while section 4 is devoted to
Fock states and to the squeezed displaced Fock states.
Finally in section 5, a brief discussion of results
and perspectives is given.

\section{The nonclassicality indicator}

The Wigner function of a state $|\psi\rangle$ defined by \cite{Wi32,HCSW84}
 \begin{eqnarray}
W_{\psi}(q,p)=\frac{1}{2\pi}\int_{-\infty}^{+\infty}dx\brkt{q-\frac{x}{2}}{\psi}
\brkt{\psi}{q+\frac{x}{2}} \exp\left(ipx\right)
\label{wdef}
\end{eqnarray}
satisfies the normalization condition
$\iint W_{\psi} (q,p) dqdp =1$.
Hence the doubled volume of the integrated negative part of the
Wigner function may be written as
\begin{eqnarray}
\delta (\psi)=\iint \left[ |W_{\psi} (q,p)| - W_{\psi} (q,p) \right] dqdp
=\iint |W_{\psi} (q,p)| dqdp -1 \ .
\label{defclass}
\end{eqnarray}
By definition, the quantity $\delta$ is equal to zero for coherent and
squeezed vacuum states, for which $W$ is non-negative.
Hence in this work we shall treat $\delta$
as a parameter characterizing the properties of the
state under consideration.
Similar quantities related to the volume of the negative part of the
Wigner function were used in  \cite{Sc99, BBCDFS02, DMWS04}
to describe the interference effects which determine the
departure from classical behaviour.

Furthermore, a closely related approach was recently advocated by
Benedict and collaborators \cite{BC99,FCM02}.
Their measure of the nonclassicality of a state $|\psi\rangle$ reads
\begin{eqnarray}
\nu(\psi)=1-\frac{I_+(\psi)-I_{-}(\psi)}{I_+(\psi)+I_{-}(\psi)}
\label{benedict}
\end{eqnarray}
where $I_+(\psi)$ and $I_{-}(\psi)$ are the moduli of the integrals over those
domains of the phase space where the Wigner function is positive and
negative, respectively. The normalization condition implies  $I_+-I_{-}=1$,
so that $\nu=2I_-/(2I_-+1)$ leads to $0\le \nu <1$.
 Using this notation we may rewrite
(\ref{defclass}) as  $\delta =I_+ +I_{-} -1=2I_-$
 and obtain a simple relation between both quantities
\begin{eqnarray}
\nu=\frac{2I_{-}}{1+2I_{-}}
=\frac{\delta}{1+\delta}
\label{deltanu}
\end{eqnarray}
with $\delta =\nu/(1-\nu)$.
It turns out that both quantities are equivalent in the sense that they
induce the same order in the space of pure states:
the relation $\delta (\psi_1) > \delta(\psi_2)$ implies
$\nu(\psi_1) > \nu(\psi_2)$. However, from a pragmatic point of view
there exists an important  difference between both quantities.

To compute explicitly the  quantity (\ref{benedict})
one faces a difficult task to identify appropriately the domains,
in which the integration has to be carried out.
On the other hand, knowing the Wigner function $W(q,p)$ of a quantum state, it
is straightforward to get its absolute value and to
evaluate numerically the integration (\ref{defclass}).

Let us emphasize again that the Hilbert space containing all pure states is huge,
so one should not expect to characterize the nonclassical features
of a quantum state just by a single scalar quantity.
 Our approach focuses on a particular issue,
whether the Wigner function is positive and may be interpreted as
 a classical probability distribution. Therefore, the proposed indicator $\delta$
should be considered as a tool complementary to these worked out earlier and
reviewed above.

\section{The Schr{\"o}dinger cat}

A quantum state, called {\sl Schr{\"o}dinger cat},
 is a coherent superposition of two coherent states
localized in two distant points of the configuration space, $\pm q_0$.
The wave function of such a state reads in the position representation
\begin{eqnarray}
\Psi(q)=\frac{N}{\sqrt{2}}\left[\phi_{+}(q)+\phi_{-}(q)\right]
\label{wftion}
\end{eqnarray}
where
\begin{eqnarray}
\phi_{\pm}(q)=\left(\frac{m\omega}{\pi\hbar}\right)^{1/4}\exp \left(
-\frac{m\omega}{2\hbar}(q \pm q_{0})^2 +i \frac{p_{0}}{\hbar}(q \pm
q_{0})\right) \ .
\label{plusmin}
\end{eqnarray}
From now on atomic units are used
$(m=\hbar=\omega=1)$.
 In other words we measure
the size of the product $pq$ in units of $\hbar$. The classical
limit $\hbar\rightarrow 0$ means the action $pq$ characteristic
of the system is many order of magnitude larger than $\hbar$.
A glance on Eq. (\ref{plusmin}) reveals that the phase, governed by
$p_0$, is of great importance in that it induces oscillations on the
wave function as can be seen in Fig.1. Note that the normalization
constant $N$ depends on the location of the centers $(q_0,p_0)$ of
both coherent states that make up the cat state. Therefore one sees that
the Wigner function may depend
not only on the distance $2q_0$ between the both states, but also
on their momentum, $p_0$.
So far, the studies on the cat states \cite{Le97}
were usually restricted to the case of standing cats, $p_0=0$.
In this letter we demonstrate that the parameter $p_0$ influences
the shape of the Wigner function, in particular,
if $q_0 \sim 1$ and both packets are not spatially separated.

Inserting $(\ref{wftion})$ into the  Wigner function (\ref{wdef})
one obtains
\begin{eqnarray}
W_{\Psi}(q,p)=W_{+}(q,p)+W_{-}(q,p)+W_{int}(q,p) \ .
\label{wcats}
\end{eqnarray}
Here
\begin{eqnarray}
W_{\pm}(q,p)=\frac{N^2}{2\pi}
\exp\left(-(q \pm q_0)^2-(p-p_0)^2\right)
\end{eqnarray}
represent two peaks of the distribution centered at the classical
phase space points $(\pm q_0,p_0)$, while
\begin{eqnarray}
W_{int}(q,p)=\frac{N^2}{\pi}\cos\left(2pq_0\right)
\exp\left(-q^2-(p-p_0)^2\right)
\end{eqnarray}
stands for the interference structure which appear between both peaks.
Normalizing (\ref{wftion}) yields
\begin{eqnarray}
N=\left(1+\cos\left(2p_0q_0\right)
\exp\left(-q_0^2\right)\right)^{-1/2} \ .
\label{normwf}
\end{eqnarray}
Making use of the formula (\ref{wcats}) for the Wigner function
of the cat state $|\Psi\rangle$ its nonclassicality parameter
\begin{eqnarray}
\delta(\Psi)=
\iint\left| W_{+}(q,p)+W_{int}(q,p)+W_{-}(q,p)\right|dqdp -1
\label{approxim}
\end{eqnarray}
may be approximated by
\begin{eqnarray}
\delta(\Psi) \approx N^2 \left[1+ \int \frac{dp} {\sqrt{\pi}}
  \left|  \cos\left(2pq_0\right)\right|
\exp\left(-(p-p_0)^2\right) \right]-1 \ .
\label{approx}
\end{eqnarray}
Strictly speaking the right hand side of equation (\ref{approx})
forms an upper bound for $\delta(\Psi)$,
which may be practically used as its fair approximation.
Because of the oscillations of the absolute value of cosine, it is
difficult to perform the integration analytically. In the special
case $q_{0}=0$, the superposition of coherent states
(\ref{wftion}) reduces to a single coherent state and correspondingly
(\ref{approx}) leads to $\delta(\Psi)=0$.

\begin{figure}
\centering
\epsfig{figure=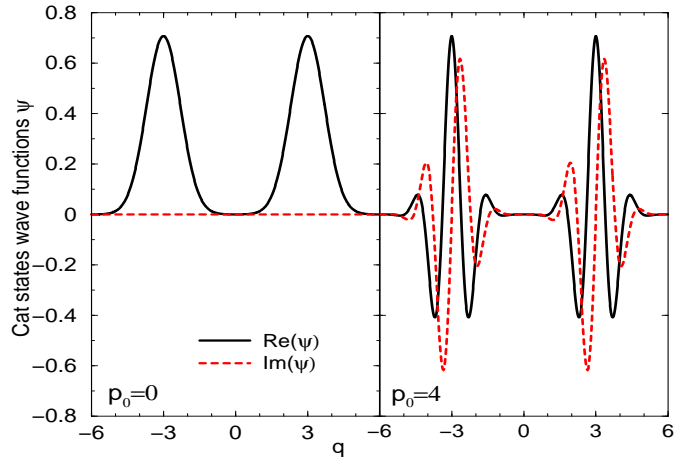,width=9cm,height=6cm}
\caption{Schr{\"o}dinger cat states wave functions plotted with $p_0=0$ (left) and with
  $p_0=4$ (right). Dashed and solid lines represent the
  imaginary and the real part of the wave function,
  respectively. Notice that the envelopes of both wave functions do coincide.}
\end{figure}

\begin{figure}
\centering
\epsfig{figure=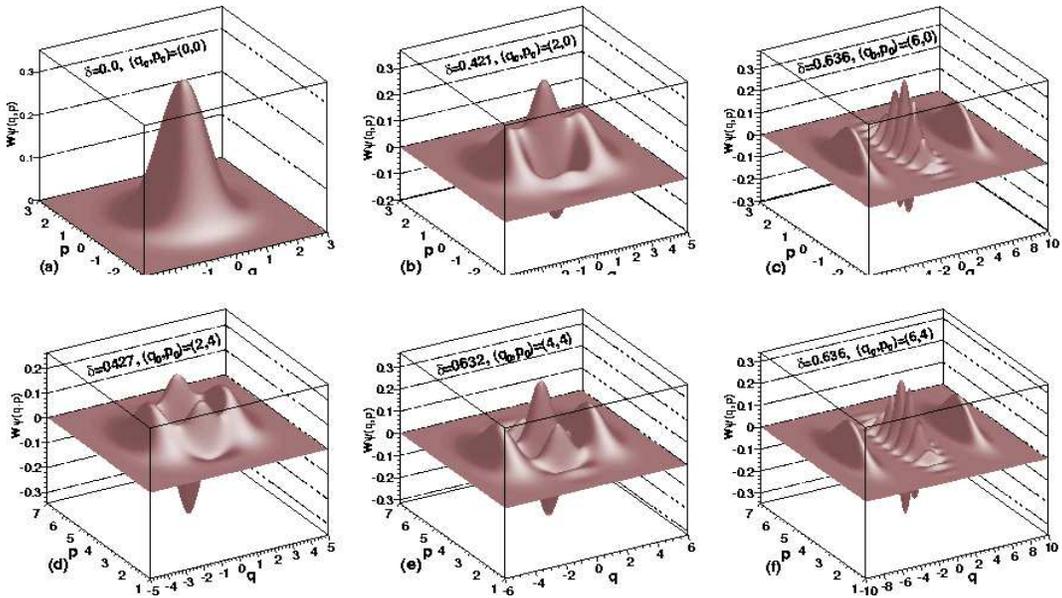,height=8cm,width=14cm}
\caption{Plots of the Wigner functions of the
  Schr{\"o}dinger cat states (\ref{wcats}). 
 Each panel is labeled by the separation distance $q_0$, 
 the momentum $p_0$ and the resulting indicator $\delta$.
Observe that for intermediate separations,  $q_0 \sim 1$,
the indicator $\delta$ changes with $p_0$. Upper row shows the
 'standing cats' ($p_0=0$) while the cats in motion ($p_0=4$) are represented in the lower row}
\end{figure}

Fig. 2 shows plots of the Wigner function of the cat states for several
values of the separation $q_0$ and the momentum $p_0$. One
clearly sees the formation of the quantum
interference structure halfway between the two humps as
the separation distance $q_0$ increases. The frequency of the interference structure
increases with the separation \cite{Le97}. 
 For intermediate separations ($0<q_0\leq 4$),
the Wigner function changes its structure with $p_0$, 
see fig.2b and 2c.
However, for a larger separation distance, $q_0>4$,
 the Wigner function for $p_0=p_1 \neq 0$
may be approximated by the Wigner function for the state
with $p_0=0$ translated by a constant vector $\Delta p=p_1$.

In the case of 'standing cats', ($p_0=0$),
 the indicator $\delta$
increases monotonically with the separation $q_0$,
and reflects presence of the interference patterns at $q=0$ - see Fig. 4k.
The growth of the nonclassicality
saturates at $q_0 \approx 4$, as the interference patterns become
practically separated from both peaks,
and the parameter $\delta$ tends to the limiting value,
$\delta_{max} \approx 0.636$. In the limit $q_0\rightarrow\infty$ the
oscillations of the cosine term in Eq. (\ref{approx}) become rapid and a
crude approximation $|\cos(q_0p_0)|\approx 1$
gives an explicit upper bound $\delta \le 2N^2-1\approx 1$.
\begin{figure}
\centering
\epsfig{figure=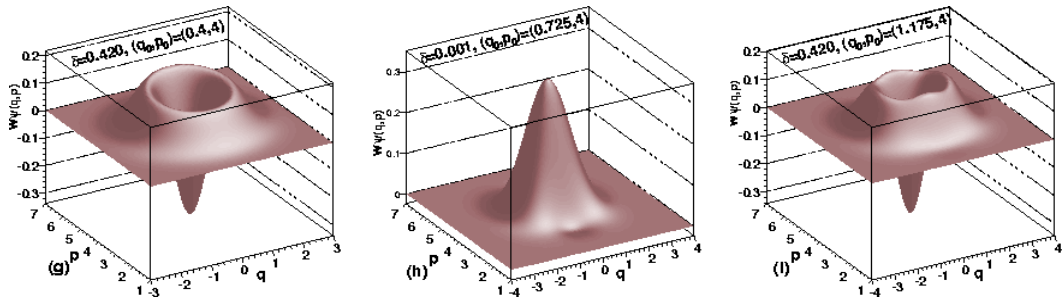,height=4cm,width=14cm}
\caption{
As in Fig. 2. The values of $q_0$
labeling each panel, correspond to the successive extrema (g,h,i)
of the indicator $\delta$ plotted in Fig. 4m as a function of $q_0$.
Wigner funtion for the 'Cats in motion' as in lower row of  Fig.2,
for the selected values of $p_0$, for which the dependence
$\delta(q_0)$ achieves its extrema.
}
\end{figure}

This picture gets more complicated for the states
with $p_o \ne 0$, in particular
for a small separation distance,  $(0<q_0\leq 4)$.
In this case, $\delta$ exhibits oscillations as
shown in Fig.4l, 4m, 4n. To shed some light on this behavior
we have chosen to plot in Fig.3 the Wigner function
for which $\delta(q_0)$ achieves extremal values.
For instance,  $\delta$ at $q_0=0.725$ (Fig. 3b)
is smaller than at $q_0=0.4$ (Fig. 3a) or $1.175$ (Fig. 3c).
This is due to the the interference structure,
which is not symmetric with respect to the reflection $p \to -p$,
in contrast to the case of cats with $p_0=0$.

As shown in Fig. 4, the frequency of oscillations
increases with $p_0$,  but the limiting value
$\delta(q_0\to \infty)$ does not depend on the initial momentum $p_0$.
This can also be demonstrated, investigating the dependence
of the quantity $\delta$ as a function of $p_0$.
As follows from Eq. (\ref{approx}), the indicator $\delta$
displays regular oscillations with
the period $p_{osc}=\pi/q_0$ -- See Fig. 5.
In other words a non--zero separation parameter $q_0$
breaks the translational invariance in momentum
and introduces a characteristic
momentum scale $p_{osc}\sim 1/q_0$.
Note that the amplitudes of the oscillations decrease
fast with $q_0$, so that 
for well separated cats with $q_0>4$
the quantity $\delta$ is practically independent on $p_0$.

\begin{figure}
\centering
\epsfig{figure=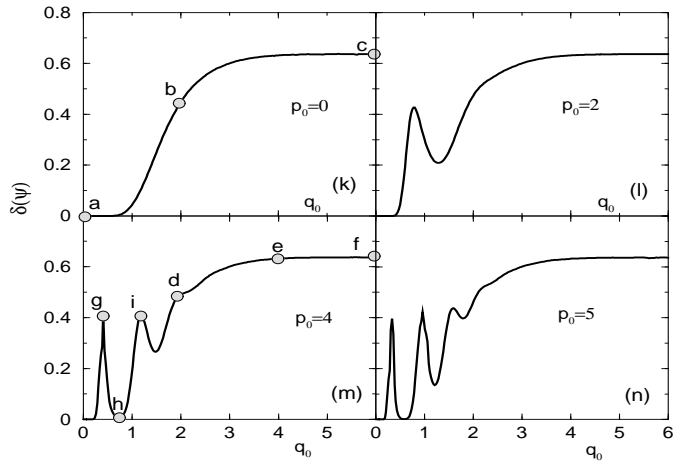,width=9cm,height=6cm}
\caption{Indicator $\delta$ of the Schr{\"o}dinger cat
 state $|\psi\rangle$ as a function of the separation distance $q_0$ and several
 values of $p_0$ as labelled on each panel. Grey dots (a-f) refer to labels
 of individual panels of Fig. 2. while grey dots (g,h,i) refer to that
 of fig.3.}
\end{figure}
\begin{figure}
\centering
\epsfig{figure=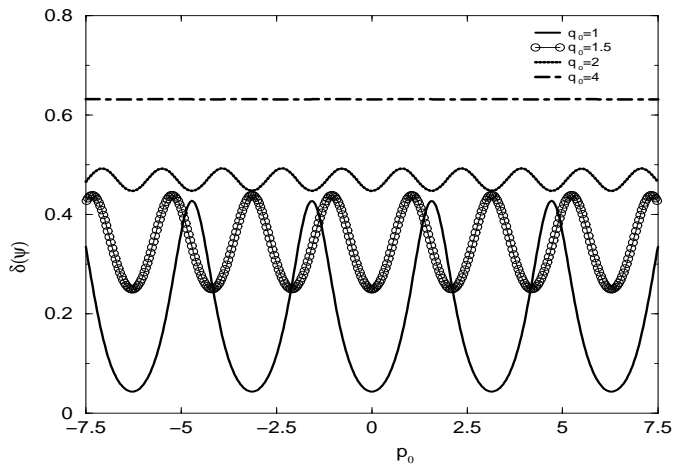,width=9cm,height=6cm}
\caption{Indicator $\delta$ of the Schr{\"o}dinger cat
 state $|\Psi\rangle$ as a function of the momentum $p_0$
for certain values of the separation $q_0$.}
\end{figure}

\section{Generalized Fock states}
Let us consider the squeezed displaced Fock state defined by
\begin{eqnarray}
\ket{\beta,\eta,n}=S(\eta)D(\beta)\ket{n} \ ,
\end{eqnarray}
where $\ket{n}$ is the original Fock state and $n=0,1,2,\dots$.
  The displacement $D(\beta)$ and the squeezed $S(\eta)$ operators are
defined by \cite{MW95,Le97}
\begin{eqnarray}
D(\beta):=\exp(\beta a^{\dagger}-\beta^* a)
{\rm \quad and  \quad}
 S(\eta):=\exp \left(\frac{1}{2}(\eta^*a^2-\eta a^{\dagger 2})\right) ,
\end{eqnarray}
where $a$ and $a^{\dagger}$ are usual photon annihilation and creation
operators, respectively.
The complex variable $\beta$
represents the magnitude and angle of the displacement.
Similarly, writing the complex number in its polar form,
$\eta=s\exp(i\phi)$, it is easy to see that the radius $s$ plays the
role of the squeezing strength while the angle $\phi$
indicates the direction of squeezing. It was shown in \cite{CG69} that the displacement operators
$D(\beta)$ form a complete set of operators. Thus any  bounded
operator $F$,
(for which the Hilbert-Schmidt norm $\parallel F\parallel=
\sqrt{{\rm Tr}(F^{\dagger}F)}$ is finite),
can be expressed in the form $F=\int f(\xi)D^{-1}(\xi){d^2\xi}/{\pi}$
in which the weight function $f(\xi)={\rm Tr}(FD(\xi))$ is unique and
square-integrable. Given that every density operator is
bounded (${\rm Tr}(\rho^{\dagger}\rho)={\rm Tr}(\rho^2)\le1$), one may write an
arbitrary density operator $\rho=\int \chi(\xi) D^{-1}(\xi) {d^2\xi}/ {\pi}$.
Here the weight function $\chi(\xi)={\rm Tr}(\rho D(\xi))$
is just the expectation value of the displacement operator commonly
known as characteristic function. The complex Fourier transform of
$\chi(\xi)$ defines the Wigner function
\begin{eqnarray}
W(\alpha)=\int \frac{d^2\xi}{\pi}\chi(\xi)\exp(\xi^*\alpha-\xi \alpha^*) \ .
\label{Wchi}
\end{eqnarray}
One may therefore express $\chi(\xi)$ in terms of the Wigner function
by performing the inverse Fourier transform as
\begin{eqnarray}
\chi(\xi)=\int\frac{d^2\alpha}{\pi}W(\alpha)\exp(\xi\alpha^*-\xi^*\alpha) \ ,
\label{chiW}
\end{eqnarray}
so that upon substitution into the density operator expression above, one gets
\begin{eqnarray}
\rho=\int \frac{d^2\alpha}{\pi} W(\alpha)T(\alpha) \ .
\label{rhoWT}
\end{eqnarray}
The  operators $T(\alpha)$ denote
\begin{eqnarray}
T(\alpha)&=&\int\frac{d^2\xi}{\pi}
\exp(\xi\alpha^*-\xi^*\alpha)D^{-1}(\xi)\nonumber\\
&=&2D(\alpha)(-1)^{a^{\dagger}a}D^{-1}(\alpha)\ ,
\end{eqnarray}
so that the Wigner function may be interpreted as a weight function for the
expansion
of the density operator in terms of the operators $T(\alpha)$ \cite{CG69}.
These operators are Hermitian, $T=T^{\dagger}$,
 and possess the same completeness properties as
the displacement operators $D(\alpha)$.
 Making use of the parity operator
$(-1)^{a^{\dagger}a}=\exp(i\pi a^{\dagger}a)$,
one finally shows that
\begin{eqnarray}
W(\alpha)=2(-1)^n {\rm Tr}(\rho D(2\alpha))
\end{eqnarray}
with $n=a^{\dagger}a$ being the photon number.

In the case of the squeezed displaced Fock states,
$\rho=|\beta,\eta,n\rangle\langle \beta,\eta,n|$, the Wigner function
becomes
\begin{eqnarray}
W_n(\alpha)=2(-1)^n\langle \beta,\eta,n|D(2\alpha)|\beta,\eta,n\rangle
\end{eqnarray}
Performing explicitly calculations of matrix elements, one obtains :
\begin{eqnarray}
W_n(\alpha)=\frac{2}{\pi}(-1)^n\exp(-2|b|^2)L_n(4|b|^2)
\label{wigner}
\end{eqnarray}
with $b=\cosh(s)(\alpha^*-\beta^*)+\exp(-i\phi)
\sinh(s)(\alpha-\beta)$\cite{kader03}. Here  $L_n$ denotes the
Laguerre polynomial of the $n$-th order.

\begin{figure}
\centering
\epsfig{figure=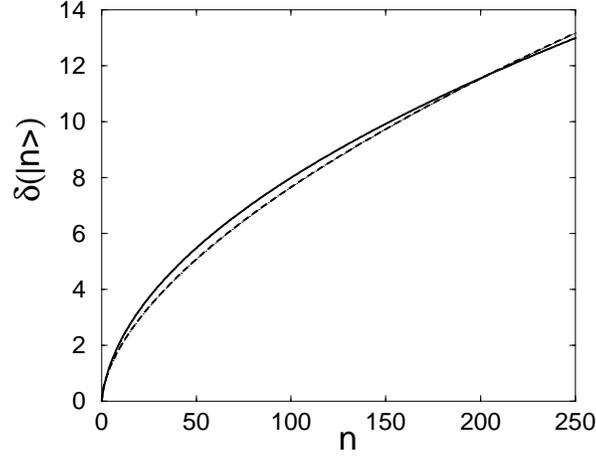,width=8cm,height=6cm}
\caption{The nonclassicality indicator $\delta(\ket{n})$ of the Fock states
  versus the quantum number $n\le 250$ (solid line). Dashed line represents
$\frac{1}{2}\sqrt{n}$ plotted for comparison. }
\end{figure}
The Wigner function (\ref{wigner}) allows us to compute
the nonclassicality parameter $\delta( \ket{\beta,\eta,n})$  for a
given displaced squeezed Fock state $|\beta,\eta,n\rangle$.
In what follows certain special cases will be investigated such as
squeezed displaced vacuum states, pure Fock states and squeezed displaced Fock
states. It will be therefore convenient to represent the
complex variable $\alpha$ by the position and momentum coordinates,
 $\alpha=\frac{1}{\sqrt{2}}(q+ip)$,
and treat likewise the displacement operator,
$\beta=\frac{1}{\sqrt{2}}(q_0+ i p_0)$.

Substituting $\beta=\eta=0$ in eq. (\ref{wigner})
yields the Wigner function for the Fock state $\ket{n}$,
\begin{eqnarray}
W_n(q,p)=\frac{(-1)^n}{\pi}\exp\left[-\left(q^2+p^2\right)\right]
L_n\left[2(q^2+p^2)\right]  \ .
\label{wfstate}
\end{eqnarray}
This allows to evaluate analytically the indicator
 $\delta(\ket{n})$, for $n=1,2,3,4$
\begin{eqnarray}
\delta(\ket{0})&=&0 \quad (vacuum) \nonumber\\
\delta(\ket{1})&=&\frac{4}{e^{1/2}}-2\approx 0.4261226\nonumber\\
\delta(\ket{2})&=&
4\left((2+\sqrt{2})e^{-1-\frac{1}{\sqrt{2}}}+(-2+\sqrt{2})
e^{-1+\frac{1}{\sqrt{2}}}\right)\approx0.72899 \\
\delta(\ket{3})& \approx &0.97667\nonumber\\
\delta(\ket{4})& \approx &1.19138 \ ,\nonumber
\end{eqnarray}
since the zeros of the Laguerre polynomials
are available up to the $4$--th order.
For larger $n$ we computed the quantity $\delta(\ket{n})$
numerically and plotted in Fig. 6.
The indicator $\delta$ grows monotonically with $n$,
as the number of zeros of the 
Laguerre polynomial $L_n$ increases with $n$.
For $n \in [1,250]$ this dependence may be aproximated by 
$\frac{1}{2}\sqrt{n}$.
Hence, the larger the quantum number $n$,
the less the Wigner function
$W_{|n\rangle}$ can be interpreted
as a classical distribution function.

Setting $n=0$ in (\ref{wigner}) one obtains a squeezed coherent state
or squeezed vacuum state.
Choosing the squeezing angle $\phi=0$,
one sees that the Wigner function is a Gaussian
centered at the displacement vector ($q_0,p_0$)
with the shape determined by the squeezing parameter $s$, 
\begin{eqnarray}
W_0(q,p)=\frac{1}{\pi}\exp\left(-e^{2s}(q-q_0)^2-\frac{1}{e^{2s}}
  \left(p-p_0\right)^2\right) \ .
\label{wsqstate}
\end{eqnarray}
In such a case the Wigner function remains everywhere non--negative for any choice
of the squeezing and displacement parameters \cite{Hu74},
 so that the nonclassicality indicator vanish,
$\delta(\ket{\beta,s,0})=0$.
Note that the displacement of any state in phase space does not change
the shape of the Wigner function, so the quantity $\delta$ is
independent of the displacement operator $D(\beta)$.

Furthermore, the
squeezing operator $S(\eta)$ influences the shape of the Wigner
function, but does not lead to a change in the volume of its negative
part. Therefore, the parameter $\delta$
does not also depend on the squeezing. As an illustration
we have chosen the squeezed ($|\alpha|=s,\phi=\pi/6$)
displaced ($\beta=0$) third photon ($n=3$) state,
$|0,s\exp(i\pi/6),3\rangle$.
The contour plots of the Wigner function of 
such a state are shown in Fig. 7 for some values of  the squeezing
parameter $s$. The indicator $\delta$ is equal to $0.9762$, irrespective
to the squeezing strength.
If squeezing is strong enough, the ring-like Wigner function 
collapses to a quasi one dimensional object with a cigar form.

The squeezed vacuum is often described
as a nonclassical state \cite{Le97}.
Since the quantity $\delta$ 
does not depend on squeezing, it should not 
be interpreted as the only
parameter which characterizes the nonclassicality.
To describe the nonclassical features of the squeezed states one may use,
for instance, the nonclassical depth \cite{Le91,LB95,MB03}.

\begin{figure}
\centering
\epsfig{figure=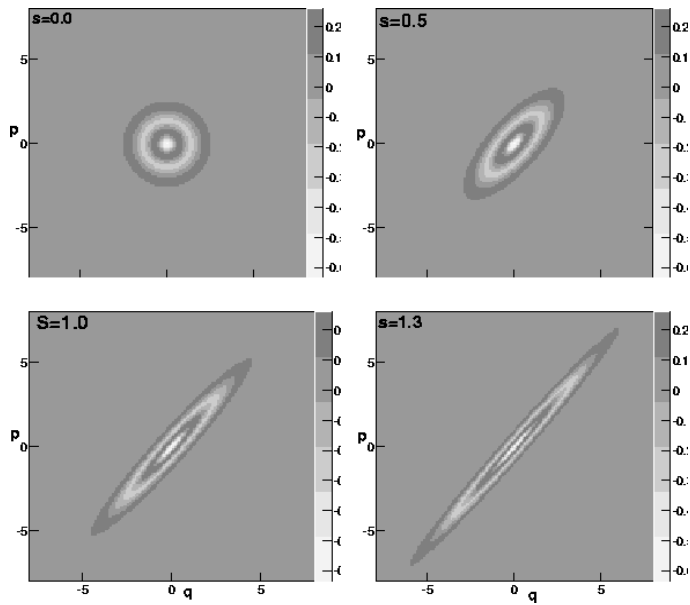, height=8cm,  width=9cm}
\caption{Contour plots of the Wigner functions of the squeezed
  Fock states $|0,s\exp(i\pi/6),3\rangle$
 labeled by the squeezing strengths $s$. 
 Irrespectively of $s$ the indicator $\delta \approx 0.97$.}
\end{figure}

\section{Concluding remarks}

In this work  we have proposed a simple indicator
of non-classicality which measures the volume of the
negative part of the Wigner function.
Although the proposed coefficient $\delta$
is a function  of the related quantity $\nu$, 
recently introduced by Benedict, Czirj{\'a}k et al. \cite{BC99,FCM02},
it is much easier to compute numerically.

The quantity (\ref{defclass})
was used to analyze exemplary quantum states,
including the Schr{\"o}dinger cat states.
The nonclassicality
$\delta$ increases with the separation between
the classical points defining the cat state.
This growth saturates, if the separation distance
is so large that the quantum interference patterns
are well isolated from both main peaks of the distributions. Moreover,
for a non--zero momentum $p_0\neq 0$, the quantity $\delta$ undergoes oscillations
until the separation distance becomes so large
that both packets are separated from the interference patterns.
Asymptotically, if the separation is large enough,
the indicator $\delta$ does not depend on $p_0$
and tends to a constant value, $\delta_{max}\approx 0.636$.

In the case of Fock states $|n\rangle$, the quantity $\delta$
equals zero for the coherent vacuum state $|0\rangle$
and grows monotonically with the quantum number $n$.
If a quantum state is displaced by the Glauber operator $D(\beta)$,
the shape of the Wigner function and the
nonclassicality parameter do not change.
Although the squeezing operator $S(\eta)$
changes the shape of the Wigner function,
our results obtained for the squeezed Fock states
show that the nonclassicality $\delta$ does not 
depend on squeezing.

The results presented in this work
were obtained for pure states of
infinite dimensional Hilbert space
with use of the standard harmonic oscillator coherent states.
It is worth to emphasize that our approach is
also suited to analyze mixed quantum states.
Furthermore, one may study the similar problem for
quantum states of a finite dimensional Hilbert space,
which was originally tackled in \cite{BC99}.
In such a case one defines the Husimi
function with the help of the $SU(2)$, spin coherent states,
while the Wigner functions may be obtained by
expanding the density matrix in the
complete basis of the rotation operators \cite{Ag81,DAS94,HW00}.
The Wigner function
for finite dimensional systems may also be defined
in alternative ways - see
\cite{Wo87,GTP88,VP90,Le95,MKI01,MCS03,Wo03} and references therein.
Studying the volume
of the negative part of the Wigner function,
defined according to any of these approaches,
one may get an interesting information
concerning the nonclassical properties of the
state analyzed. For instance  some recent
attempts \cite{Ga04,DMWS04,Be04}
try to link the negativity of the Wigner function
with the entanglement of analyzed
quantum states defined on a composed Hilbert space,
or with the violation of the Bell inequalities.

\section{Acknowledgment}

We are indebted to I. Bia{\l}ynicki--Birula, I. Bengtsson,
J. Burgd{\"o}rfer, A.R.R. Carvalho, E. Galv{\~a}o,
A. Miranowicz, A.M. Ozorio de Almeida,
J. M. Rost, K. Rz{\c a}{\.z}ewski and M.S. Santhanam
for fruitful discussions, comments and remarks.
 We also would like to thank
P.W. Schleich for helpful correspondence and
several remarks that allowed us to improve the manuscript.
Financial support by
Polish Ministry of Scientific Research under
the grant No PBZ-Min-008/P03/2003 and the VW grant
'Entanglement measures and the influence of noise'
is gratefully acknowledged.
AK gratefully acknowledges the financial support by Alexander von Humboldt
(AvH) Foundation/Bonn-Germany, under the grant of Research fellowship
No.IV.4-KAM 1068533 STP.

\end{document}